\documentclass[twocolumn,prb]{revtex4}

\ifx\pdfoutput\undefined
\usepackage{graphicx}
\else
\usepackage[pdftex]{graphicx}
\usepackage{epstopdf}
\fi

\usepackage[center]{subfigure}
\usepackage{amsmath}

\begin{document}

\title{Steered Transition Path Sampling}

\author{Nicholas Guttenberg and Aaron R. Dinner}
\affiliation{James Franck Institute, The University of Chicago, Chicago, IL 60637}
\email{dinner@uchicago.edu}

\author{Jonathan Weare}
\affiliation{Department of Mathematics, The University of Chicago, Chicago, IL 60637}
\email{weare@math.uchicago.edu}

\begin{abstract}
We introduce a path sampling method for obtaining statistical properties of an
arbitrary stochastic dynamics.  The method works by decomposing a trajectory in
time, estimating the probability of satisfying a progress constraint, modifying
the dynamics based on that probability, and then reweighting to calculate
averages.  Because the progress constraint can be formulated in terms of 
occurrences of events within time intervals, the method is particularly
well suited for controlling the sampling of currents of dynamic events.  We
demonstrate the method for calculating transition probabilities in barrier
crossing problems and survival probabilities in strongly diffusive systems with
absorbing states, which are difficult to treat by shooting.  We discuss the 
relation of the algorithm to other methods.
\end{abstract}

\maketitle

\section{Introduction}

Over the past decade and a half, there have been dramatic advances in the
ability to sample dynamical events that are rare with respect to the time step
used for numerical integration of models \cite{Dellago2009,Dickson2010_review}.
The core idea is that it is necessary to account for the statistics of
trajectories.  The earliest practical such method and still the most widely
used is transition path sampling (TPS), a Monte Carlo procedure
\cite{Dellago1998,Frenkel2002,Dellago2009}.  Given a trajectory that satisfies
constraints of interest, a portion of the trajectory is modified (the trial
move) and then accepted or rejected so as to sample a desired path ensemble.  

The fact that TPS is a Monte Carlo procedure suggests that one has flexibility
with respect to both the move set and the acceptance criterion, so long as
detailed balance in the path space is satisfied.  However, the only moves that
are routinely employed are ``shooting'', in which a time point is selected, and
then all steps from that point to one or both ends are replaced by integrating
the original equations of motion, and ``shifting'', reptation of the
trajectory, again based on the original dynamics; in these cases, the
acceptance criterion that leads to the physically weighted transition path
ensemble reduces to a simple check for whether the defining constraints are
satisfied \cite{Dellago1998b}.

Given that it is often advantageous to sample a non-physically weighted
ensemble in Monte Carlo simulations (e.g., as in umbrella sampling for enhanced
exploration of selected order parameters \cite{Torrie1977,Frenkel2002}), it is
striking that this strategy has limited representation in the path sampling
literature.  Dynamic importance sampling (DIMS)
\cite{Zuckerman2001,Perilla2011} is built around the idea that one can bias the
dynamics and reweight the paths.  The problem is that the path probability is a
product over the step probabilities and thus depends exponentially on
trajectory length.  Consequently, quantitative averages (e.g., rates) fail to
converge in practice \cite{Perilla2011}.  That said, the overlap between path
ensembles with different extents of bias is sufficient for annealing to
generate initial unbiased paths for TPS efficiently \cite{Hu2006b}, and DIMS
appears to yield qualitatively reasonable paths that can provide mechanistic
insight \cite{Perilla2011}.

One successful quantitative example of path reweighting is in the calculation
of free energy differences using Jarzynski's nonequilibrium work theorem
\cite{Jarzynski1997}.  In that case, the average of interest,
$\langle\exp[-\beta W]\rangle$, where $\beta$ is the inverse temperature
and $W$ is the work, is dominated by the trajectories in which the system
spontaneously does work on its surroundings (negative work).  $W$ can be
irreversible, but when the system is driven hard, the negative-work
trajectories are rare and the distribution of work is skewed.  
Sun \cite{Sun2003b} showed that sampling a work-weighted path ensemble with TPS
yielded free energies of a desired precision with significantly less
computational effort than physically weighted nonequilibrium simulations.

More recently, Hernandez and co-workers \cite{Ozer2010} showed that they could
accelerate the convergence of free energies from nonequilibrium work by
decomposing the process of interest (in their case, force-induced protein
unfolding) into a series of physically weighted shorter ones connecting
configurations along a guiding path.  The advantage is that the work
distributions of the shorter processes are closer to Gaussian, so that the
factorized contributions converge much faster than $\langle\exp[-\beta
W]\rangle$.  This is an example of the general rule that decomposing a
small probability into a product of probabilities near unity speeds the
overall calculation; this idea is central to many methods for estimating rates:
transition interface sampling (TIS) \cite{vanErp2003}, milestoning
\cite{Faradjian2004,West2007}, forward flux sampling (FFS)
\cite{Allen2005,Allen2009}, the weighted ensemble method (WE)
\cite{Huber1996,Bhatt2010}, nonequilibrium umbrella sampling (NEUS)
\cite{Warmflash2007,Dickson2009a,Dickson2009b,Dickson2010,Dickson2011},
and boxed molecular dynamics (BXD) \cite{Glowacki2009}.

Here we combine path reweighting with decomposition to introduce a general
path sampling algorithm that affords excellent control over arbitrary events
and efficient calculation of quantitative averages:  steered transition
path sampling (to which we refer with the acronym STePS for mnemonic reasons).
In particular, the method allows accumulation of rare numbers of events that
are individually common for control of currents in a space of order parameters,
as previously sampled by TPS \cite{Merolle2005,Hedges2009} and methods for
calculating large deviation functions \cite{Giardina2006,Tchernookov2010b}.
The fact that all integration is forward in time allows treatment of
microscopically irreversible systems and the study of relaxation from arbitrary
initial conditions.  A final feature is that, like TPS, whole trajectories are
obtained, facilitating calculation of arbitrary multi-time correlation
functions.  We demonstrate the method on a series of examples.  Then we discuss
its relation to other methods.

\section{Algorithm}

An event is rare when many attempts must be made by a system before achieving a
successful realization.  A typical such situation in molecular systems is that
a system is trapped in a region of phase space, with relatively few
trajectories leading out of it.  Shooting \cite{Dellago1998b}
enables efficient harvest of these trajectories for systems with
microscopically reversible dynamics by always integrating away from a
bottleneck and toward a stable state (i.e., downhill in free energy).  When
there are multiple local minima, however, the path must contain some uphill
segments, which can drastically lower the acceptance rate.  In some cases, this
issue can be addressed by manually breaking the process of interest into a
small number of steps, each of which involves only a single bottleneck (e.g.,
\cite{Radhakrishnan2004,Hu2008}), but doing so requires considerable insight into a
system.

We propose an algorithm that naturally finds trajectories that link multiple
elementary steps to achieve an overall dynamics of interest.  The idea
is that we iteratively generate many short segments of unbiased dynamics and
select among them for those that satisfy a constraint on the progress of the
rare event in question (e.g., that an order parameter must increase in value).
As mentioned in the Introduction, this is similar in spirit to forward flux
sampling and related methods
\cite{vanErp2003,Faradjian2004,West2007,Allen2005,Allen2009,Huber1996,Bhatt2010,%
Warmflash2007,Dickson2009a,Dickson2009b,Dickson2010,Dickson2011,Glowacki2009},
but the present case does not require
definition of interfaces in the space of variables that describe the system.
More importantly, we select among the segments at each interval in time with
non-physical weighting and then correct for this bias in calculating averages.
To this end, we use the trajectory segments to estimate the probability $P$
that the progress constraint is satisfied over the interval.  If $P$ is greater
than a threshold $Q$, we select among the segments with uniform probability
(i.e., physical weighting).  On the other hand, if $P<Q$, we select the forward
direction with probability $Q$, 
and the path probability must be modified by a reweighting factor.  We
accumulate a product of such factors as we build up the full trajectory.

The precise steps of the algorithm are as follows.  

\begin{enumerate}
\item{
Initialize the system as in a standard dynamics simulation; 
initialize the overall path-probability reweighting factor for the $W$ to one.
}
\item{
\label{step:burst}
Generate a set number of stochastic dynamics segments of length $\Delta$.
Denote the total number of segments at this time interval by $N$ and the
probability that a segment satisfies the constraint by $P$.  Initially, we
increase $N$ in batches of fixed size $M$ until at least one segment satisfies the progress
constraint; as the simulation progresses, we use batches of size $2/\langle P(\phi)\rangle$
where $\langle P(\phi)\rangle$ is a running average value of $P$ at order
parameter value $\phi$.  Given the $N$ segments, estimate $P$ as the fraction
that satisfy the constraint.
}
\item{
Select to satisfy the constraint with probability $R=\max(Q,P)$.  
\begin{enumerate}
    \item{If the constraint is met, set the system variables to be
    those at the end of the last such segment.
    Multiply $W$ by $P/R$.
    }
    \item{Otherwise, set the system variables to be those at the end of the last 
    segment that did not satisfy the constraint.   Multiply $W$ by $(1-P)/(1-R)$.
    }
\end{enumerate}
}
\item{
Add $\Delta$ to the time $t$.  If $t$ is less than the desired trajectory
length $\tau$, go to step \ref{step:burst}.
}
\item{
Weight contributions of this path to averages by $W$.
}
\end{enumerate}
This procedure can be repeated with different random number generator seeds 
to harvest as many trajectories as desired.  

There are a number of parameters that affect the efficiency of the algorithm.
The computational cost of a STePS simulation is directly proportional to the
number of segments used to estimate $P$, so one wants to use as few as possible
while still obtaining trajectories that satisfy the constraint; potential error
in the procedure used to estimate $P$ is discussed in Appendix \ref{secn:bias}.
In addition, the efficiency depends on $Q$, which sets the extent of bias, 
and $\Delta$, the length of the dynamics segments. 
We consider each of these below.  

\subsection{Choosing the bias threshold ($Q$)}
\label{secn:choosingq}

One can in principle use an arbitrarily complex $Q$ so long as it is known.
While we explored an adaptive $Q$ that varied with the order parameter (data
not shown), we found that the performance was not markedly better than that
obtained with a fixed value of $Q$.  How should we choose this value?  We want
the bias to drive the system forward, but not so strongly that it cannot fall
back to search for alternative routes when it becomes stuck; forcing the system
through bottlenecks can result in paths that have very low weights.  

To obtain a more quantitative perspective, we consider the effect of $Q$
explicitly for a simple problem.  Specifically, the weight in the simple case in
which the value of $P$ is nonrandom, constant, and less than $Q$ is
\begin{equation} W =
\left(\frac{P}{Q}\right)^m\left(\frac{1-P}{1-Q}\right)^{n-m} \end{equation}
where $n=\tau/\Delta$ is the total number of integration intervals, and $m$ is
the number of intervals in which progress was made. 
Ultimately, it is the variance in an estimate of a quantity of interest that
matters, not the variance in $W$.  With this caveat, we see that $Q$ values
close to one are likely to be problematic because they cause the weight
distribution to have large variance.  

Furthermore, for a given value of $Q$, the variance of the weights
increases as the number of intervals $n$ increases, resulting in a degradation
in the statistical quality of STePS estimates for long trajectories.  In other
words, it is important to focus the bias on the transition path part of a
trajectory so as to avoid accumulating large reweighting factors for
trajectory segments that simply fluctuate within a stable state.  

Relatedly,
the considerations in  Appendix \ref{secn:drift} suggest that we should choose
$Q-P = 1/\sqrt{n}$, and it is straightforward to show that this scaling leads $S
\equiv \ln W$ to tend to a Gaussian random variable for large $n$, meaning that
the weights remain well-behaved.  This scaling also suggests that the choices
of $Q$ and $\Delta$ (through $n$) are coupled, so that it is important to bias
less in the individual intervals as we increase their number.

\subsection{Choosing the segment length ($\Delta$)}

We can examine the dependence on the number of intervals, and thus the segment
length, for a more realistic choice of dynamics than above.  To this end,
we consider a one-dimensional system with overdamped Langevin dynamics
\begin{equation}
\label{eq:overdamped}
\dot{x} = F + \sqrt{2T}\dot{B}, 
\end{equation}
where $x$ is the position, $F$ is a systematic force, $T$ is the temperature,
and $\dot{B}$ is a Gaussian white noise.  For the moment we assume that
$F$ is constant.
Let the progress constraint be that $x$ increases over the course of an interval
(i.e., $x(t+\Delta) > x(t)$ for $t$ values that are integral multiples of
$\Delta$).  

The distribution of positions at the end of an integration interval is
\begin{equation}
\rho(x(t+\Delta)) = 
\frac{\exp\left[-(x(t+\Delta) - x(t) - F \Delta)^2/4 T \Delta\right]}{\sqrt{4 \pi T \Delta}}
\end{equation}
and the probability of satisfying the constraint is
\begin{align}
P &= \int_{x(t)}^{\infty} \rho(x(t+\Delta)) dx(t+\Delta) \notag \\
  &= \frac{1}{2} \left[ 1 + \textrm{erf}\left(\frac{F}{2} \sqrt{\frac{\Delta}{T}}
\right) \right].
\end{align}
We see that the probability that the progress constraint is satisfied depends
 on the length of the segments.  For the overdamped integrator
considered, as $\tau\rightarrow 0$, $P\rightarrow 1/2$, so progress is
guaranteed
if $\Delta$ is sufficiently small. This is not true in general. For
inertial dynamics, the shortest interval that should be used is on the order of
the correlation
time of the order parameter.  

In practice, when transitioning over
steep potential barriers, short intervals are needed to prevent repeatedly
simulating lengthy segments that ultimately fail to satisfy the progress
constraint.  
We analyze the continuum limit (small $\Delta$) of STePS in more detail in
Appendix \ref{secn:drift}. 
More generally, there are two competing contributions to the
variance of an estimate in a sampled quantity for a fixed overall computational
cost.  One is the number of segments that need to be generated to estimate $P$,
which scales as $1/P(\Delta)$; this becomes large on uphill sections of the
trajectory when $\Delta$ is large.  The other contribution comes from the fact
that the path weights become very broadly distributed as the number of
integration intervals increases.  While we argued this qualitative trend from
the
standpoint of a specific model in the previous section, it generally holds.  
Below, we vary $\Delta$ to illustrate how it affects convergence in practice.

\section{Examples}

In this section, we demonstrate STePS on a series of illustrative
problems.  Because the quantity of interest is often small (e.g., a transition
probability), we measure the average time to achieve a desired accuracy in the
logarithm of an estimate:
\begin{equation}
\epsilon = \left|\ln\left(\frac{\textrm{estimate}}{\textrm{reference}}\right)\right|.
\end{equation}
We use the deviation in the logarithm of the
estimate rather than the estimate itself because it better distinguishes an
estimate that is of the correct order of magnitude from one that is simply
zero owing to a failure to observe any events.

\subsection{Barrier crossing}

\begin{figure}
\includegraphics[width=\columnwidth]{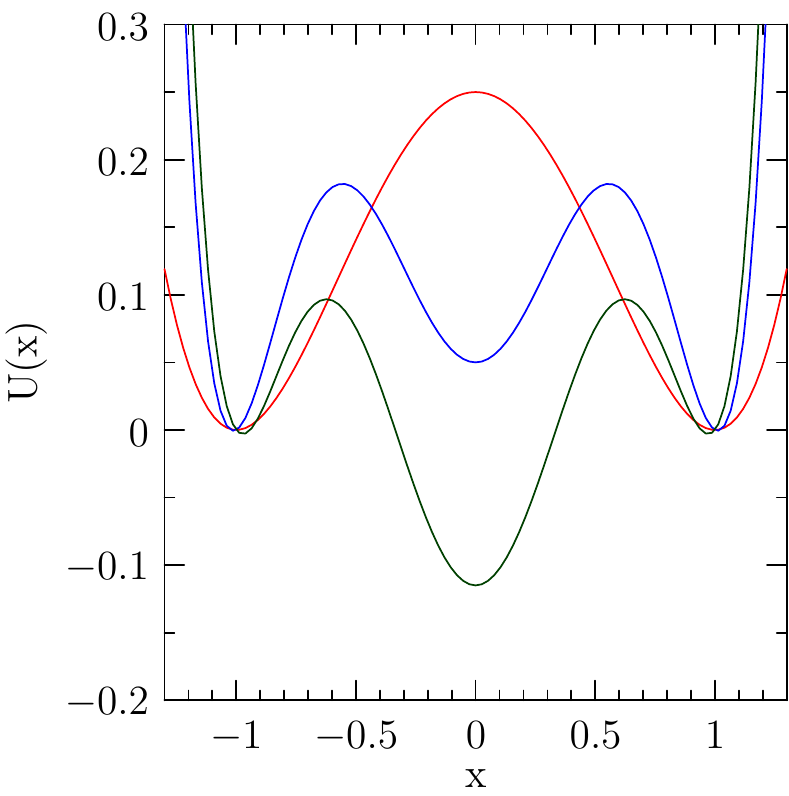}
\caption{Double and triple well potentials.
In the case of the double well, the barrier has a height of $0.25$.
For the triple well, we perform simulations for $\alpha=0.05$, $\beta=1$
and $\alpha=-0.1$, $\beta=1.15$.  In the former case, the primary barrier has
height $0.18$ and the secondary barrier has height $0.13$. In the latter case
the primary barrier has height $0.10$ and the secondary barrier has height
$0.21$.  All barriers are in arbitrary energy units.
} \label{Wells}
\end{figure}

\begin{figure} 
\includegraphics[width=\columnwidth]{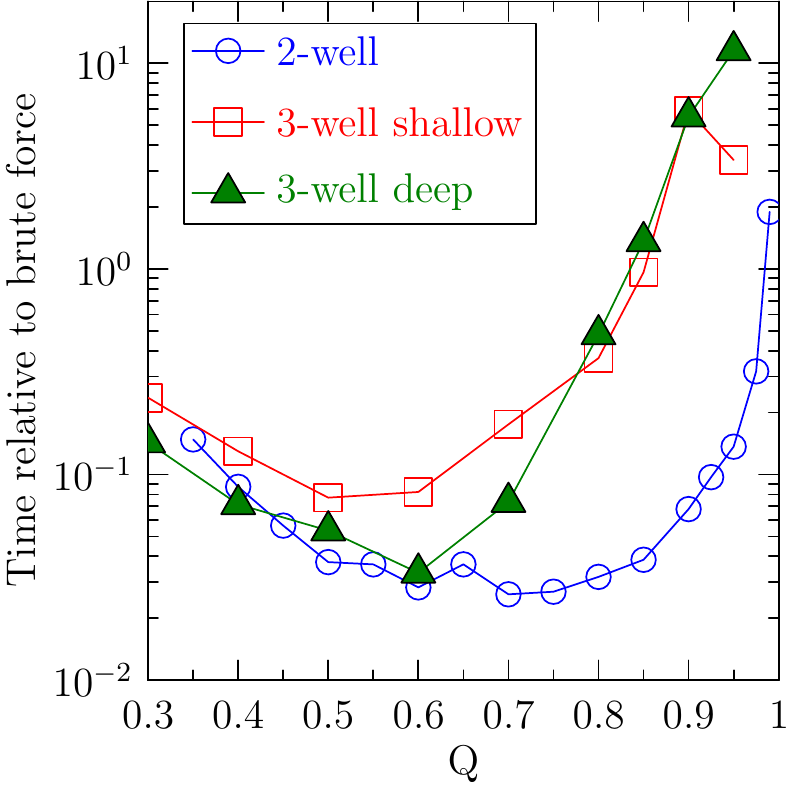}
\caption{Performance for the double and triple wells.  For the results shown,
the temperature is $T=0.02$, the time step is $5 \times 10^{-3}$, the path
length is $\tau = 20$, and the segment length is $\Delta = 0.5$. 
The initial batch size is $M=10$, and subsequent batch sizes are determined by
sampling over the order parameter $\phi=x$.
} \label{WellsCost}
\end{figure}

We first test STePS on the classic rare event problem, a barrier
crossing.  We consider both double and triple well potentials: 
\begin{eqnarray}
U(x) &=& \frac{1}{4} x^4 - \frac{1}{2} x^2\\
U(x) &=& \beta (x^6 - 2 x^4 + (1-\alpha) x^2).
\end{eqnarray}
The parameter $\alpha$ tunes the significance of the intermediate in the triple well:
$\alpha>0$ ($\alpha<0$) causes the inner well to be more (less) shallow than
the outer ones.  The parameter $\beta$ controls the well height. We use this to maintain
a roughly constant crossing probability between the two triple well cases that we examine so
as to obtain the same computational difficulty (for brute force) in both cases. 
The specific potentials that we use are displayed in
Fig.~\ref{Wells}.  We integrate the position with overdamped Langevin dynamics
(Eq.\ \eqref{eq:overdamped} with $F = -U'(x)$). 
We seek to determine the probability that a path that starts at $x =
-1$ is at $x > 1$ at a specified time $\tau$.  We obtain the reference values by
brute force; they are
$(2.9 \pm 0.3) \times 10^{-5}$, $(5.1 \pm 0.6) \times 10^{-6}$, and $(5.6 \pm 0.7) \times 10^{-6}$ 
for the double well, deep triple well, and shallow triple well, 
respectively.  
We then plot the relative time that is
required to achieve an error in the logarithm of $\epsilon=2$ in Fig.~\ref{WellsCost}.
We observe a speed-up of 10 to 30 fold for a
relatively wide range of $Q$ values, showing that the algorithm is not very
sensitive to this parameter.  This modest acceleration is consistent with the
relative ease of the problem for the selected potentials. While STePS can be
used to accelerate barrier crossings, there are many other rare event methods
for this purpose.  We show in the next section that there
is another important class of problems at which STePS truly excels.

\subsection{Absorbing Channel}

\begin{figure*}
\includegraphics[width=\columnwidth]{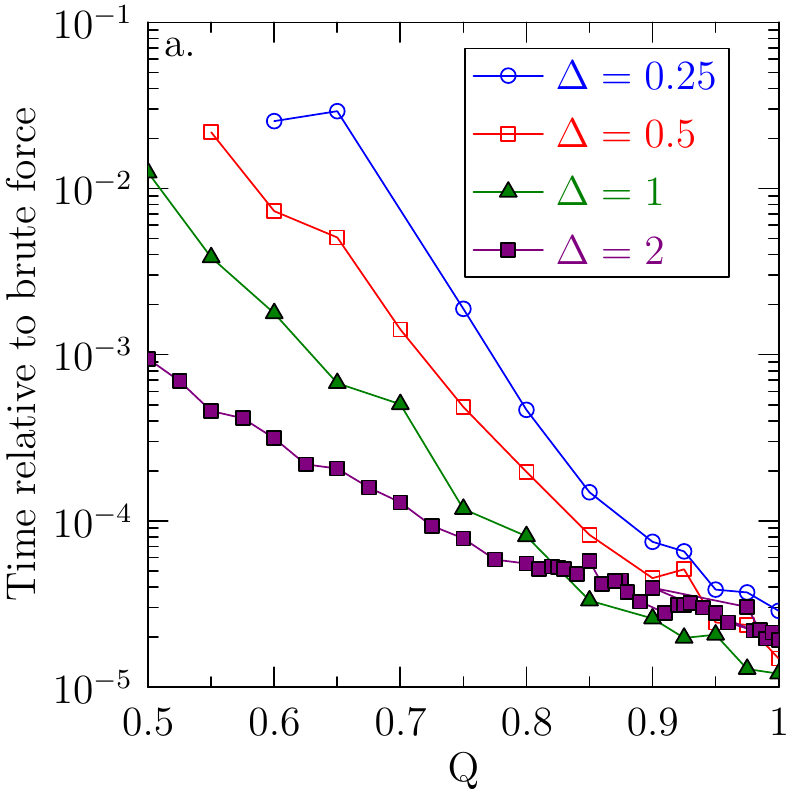}
\includegraphics[width=\columnwidth]{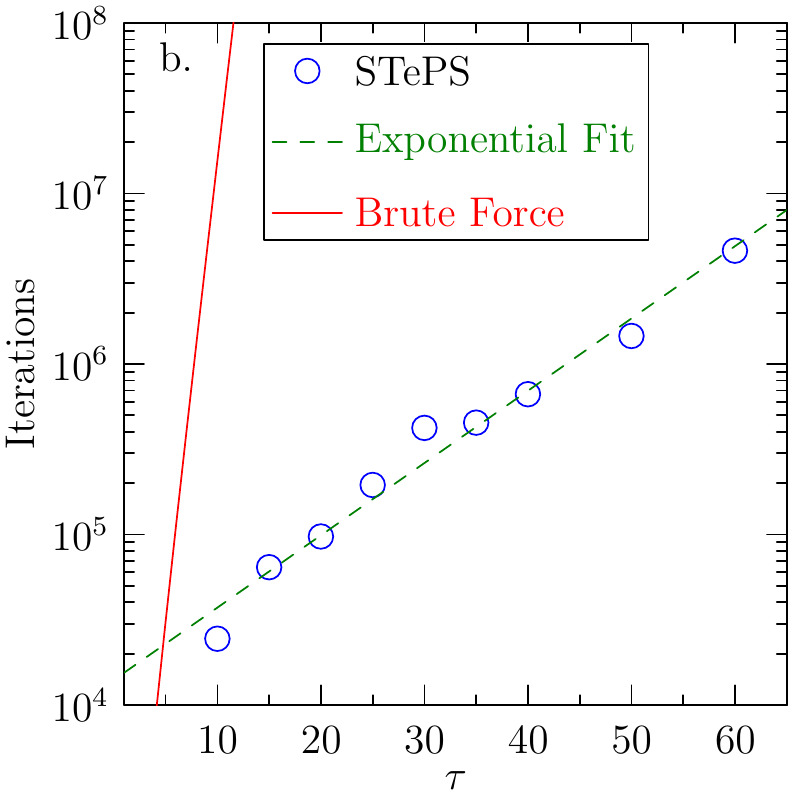}
\caption{
Performance for the absorbing channel. The initial batch size is $M=10$, with order parameter $\phi=x$ used
for future batch size estimates. All data is for the time required to achieve an error in the logarithm 
of the survival probability $\epsilon=2$. (a) Time relative to brute
force for $\tau=15$; $Q$, and $\Delta$ as indicated.  (b) 
Absolute time to convergence as a function of $\tau$; $\Delta = 1$.  The brute
force curve is expected to continue exponentially (linearly as plotted) to
$10^{22}$ iterations at $\tau=40$ compared with $10^7$ for STePS. }
\label{BrownianFigs}
\end{figure*}

\begin{figure}[t]
\includegraphics[width=\columnwidth]{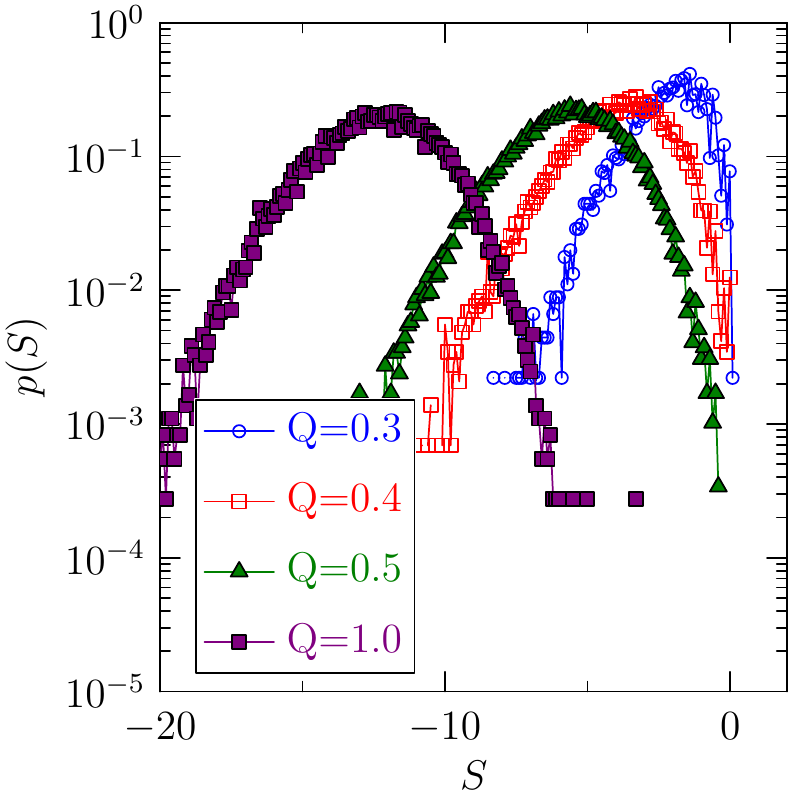}
\caption{Distribution of weights for successful
paths in the absorbing channel with $\tau=10$ for $Q$ as indicated. 
} \label{WDistFig}
\end{figure}

While STePS can boost trajectories over 
barriers as in the double well above, it performs best in situations where the
process under observation is rare because it involves an unusual number of
individually common events rather than a single bottleneck. This suggests that
one should look to apply STePS in cases where a particular current is desired,
such that at least a certain number of transitions should occur per unit time.
By the same token, STePS may do well in cases where one wants to investigate
dynamics that avoid an event---for instance, in systems with absorbing states. 

We look at one example to assess the performance of the algorithm in such
cases: Brownian motion (Eq.\ \eqref{eq:overdamped} with $F = 0$)
in a channel with absorbing walls at $|x|=1$.  The statistic that we seek
to compute is the probability $p$ that the trajectory obeys $x \in (-1,1)$ for all
$t \in [0,\tau]$ given $x(0) = 0$. To enable direct comparison to brute
force, we use an integration time step of $5\times 10^{-3}$, a temperature
$T=0.5$, and two choices of total times: $\tau=10$ and $15$. These choices
result in brute force reference probabilities of $p=1.4\times 10^{-5}$ and
$p=6.5\times10^{-8}$, respectively. 

For STePS, we define the progress constraint to be that the trajectory does not
touch the walls. We determine the relative time required to achieve an error in
the logarithm of $\epsilon=2$ for various biases $Q$ and segment lengths
$\Delta$ (Figs.\ \ref{BrownianFigs}a,b). We find optimum values of
$\Delta\approx 1$ and $Q=1$. The fact that there is no divergence as $Q\rightarrow1$ is
because all trajectories containing paths that touch the walls are not included in
the final measure. As such, even if the weights of these trajectories are
much greater than one or even divergently large, their final contribution is 
zero, and the estimate of interest is unaffected.

This system is only slightly more complex than that discussed in Section
\ref{secn:choosingq}.
As such, we look at the distribution of $S\equiv\ln W$ in these
simulations (Fig.\ \ref{WDistFig}).  We find that near the maximum of the
distribution it behaves as a Gaussian in $S$, but the low-weight tail is
exponential, extending as  $Q$ increases.

The longer we attempt to sustain the trajectory and keep it from touching the
walls, the larger the STePS speed-up relative to brute force. We cannot successfully
perform brute force simulations for $\tau>15$, however. To estimate the 
scaling of STePS for larger $\tau$ values, we compare STePS to itself. We use a single
STePS simulation of order $200$ times longer than we eventually need to satisfy the
$\epsilon=2$ bound to measure the survival probability for values of $\tau$ up to
$60$. We then use this measured probability to evaluate the convergence time of the
algorithm. 

The convergence time required by STePS is plotted alongside the estimated
convergence time for brute force (which scales as $p^{-1}$) in Fig.\
\ref{BrownianFigs}c. The difficulty of the problem is roughly exponential in
$\tau$: $C \propto \exp(\lambda \tau)$. For brute force, $\lambda = 1.10$ while
for STePS, we find that a fit of this form to the simulation time gives us
$\lambda = 0.098$, more than a 10 fold difference. As such, the 
computational advantage of STePS over brute force grows rapidly with
trajectory length.

\subsection{Adatom Diffusion}

Observing that the algorithm performs significantly better for cases in which we wish
 to obtain (or avoid) a large number of relatively common events rather than to force
  a single rare event to occur, we turn to a more complex problem in this class. We
   consider two adatoms of a particular type (red) diffusing on a crystal surface. There
    is some density of adatoms of a different type (blue) that decorate the surface and
     also diffuse. The problem that we seek to solve is to determine the probability
      that one red adatom comes into contact with the other before a blue adatom.

We simulate the dynamics with a kinetic Monte Carlo procedure on a periodic 64$\times$64 lattice. 
 In this paper we only consider the case in which all adatoms diffuse freely without interaction, though in general one can apply the same approach to cases in which
neighboring adatoms form bonds. Because all adatoms have the same probability of moving in the absence of bonds, in each iteration, we simply choose a random 
adatom, followed by a random direction to move it. If the generated move would cause it to enter an occupied cell, we simply
ignore the move. We use the total number of moves attempted as a proxy for time.

For STePS, we use an order parameter similar to that for the absorbing channel.
We consider a segment to fail to make progress if a red adatom comes into
contact with (i.e., is nearest neighbor to) a blue adatom. In the absorbing
channel simulations above, we kept track of the local success probability $P(\phi)$ by using the
distance from the walls, but, for the present problem, there is no
simple function that correlates with the probability of generating a successful
segment. As such, we instead use 
$2/r$ as the sampling batch size, where $r$ is the fraction of
moves of the red adatoms that do not put them in contact with a blue adatom.
In practice,
we estimate $r$ only at the start of a simulation.  This is ad hoc as $r$
changes during the simulation, and it does not account for the motion of
distant adatoms.  However, it incorporates the main effects of variations in
density in a straightforward manner.

In general, the segment length used should be chosen to scale with the average time between attempted contacts of the red adatoms with other
adatoms (the ``mean free time'') measured in terms of attempted particle moves (iterations). This roughly preserves
the success probability as other simulation parameters such as the density or grid
size change, as one is keeping the number of ``brushes with danger'' (potential collisions)
constant per segment. However, for simplicity, we use a fixed segment length of 1000 iterations for the simulations that we present
as it does not make a large difference in the results for the parameters that we examine. 

\begin{figure}
\includegraphics[width=\columnwidth]{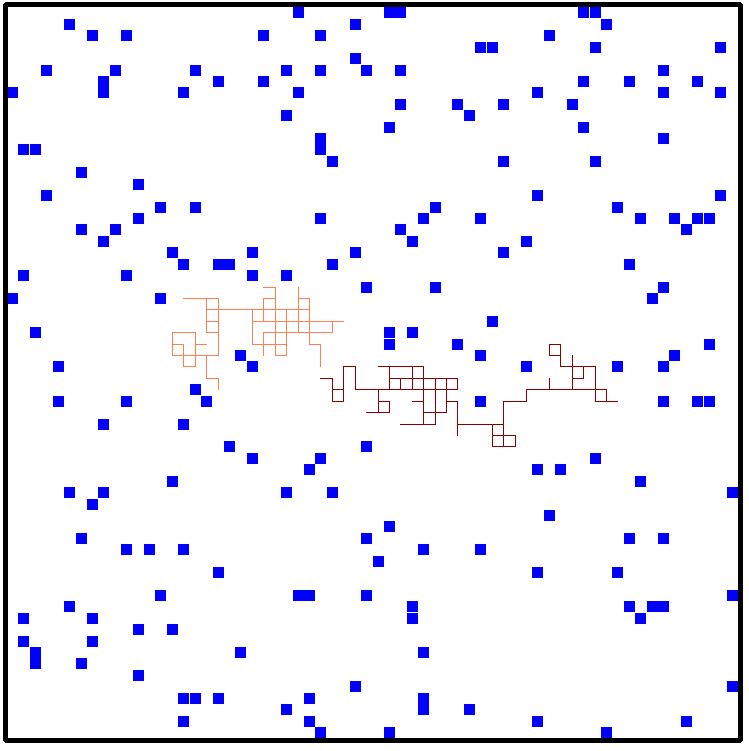}
\caption{Example trajectory of two red adatoms coming into contact with each other prior to blue adatoms at $\rho=0.05$. For the purposes of visualization, the blue adatoms are fixed in this simulation.  The two red adatom paths are distinguished by lighter and darker colors. 
}
\label{AdatomTrajectory}
\end{figure}

We show a projection of the highest weight trajectory observed over an extended
run of the simulation at fractional coverage $\rho=0.05$ in
Fig.~\ref{AdatomTrajectory}.  In this simulation, we fix the positions of the
blue adatoms for visualization purposes.  An animation of a similar simulation
with the blue adatoms allowed to move is available 
\cite{snote}. 
Note that since the red adatoms must avoid being adjacent to the blue adatoms,
each blue adatom blocks a total of five grid cells, so the percolation
transition is close to $\rho=0.1$. The diffusion of the blue adatoms softens
this threshold, as even at very high densities the blue adatoms can move to
form a path for the red adatoms.

We now examine in detail a system with fractional coverage of adatoms
$\rho=0.03$. This is the limit of density that we can reasonably
calculate via brute force.  Our initial condition consists of randomly placed
blue adatoms, with two red adatoms $32$ grid spacings apart parallel to an
axis. We simulate until a red atom makes contact with either the other red atom
or a blue adatom.  By brute force, we find that the probability that the red
adatoms contact each other first is $2\times10^{-7}$. The average time
before a collision between a red adatom and any other sort of adatom occurs is about $640$ iterations. 
The length of successful paths is much longer:  about $11000$ iterations.
For the optimal value of $Q$, around $0.9$, we obtain a factor of $20$ speed-up
compared to brute force. 

\begin{figure}[t]
\includegraphics[width=\columnwidth]{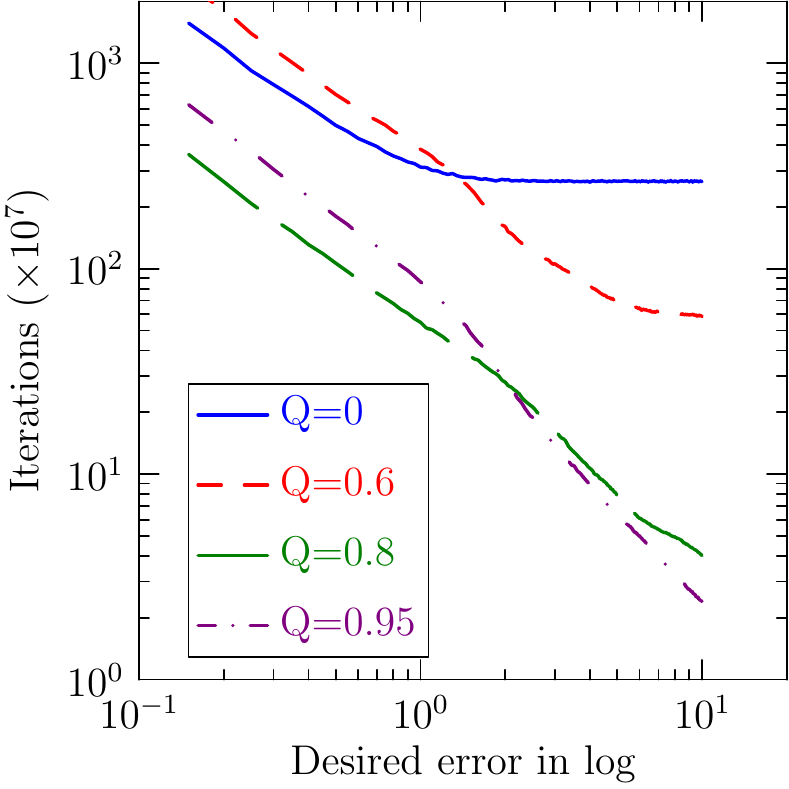}
\caption{Time required for convergence of the probability that the red adatoms contact each other at $\rho=0.03$.  The curves are obtained by the resampling procedure described in the text.  The $Q=0$ curve corresponds to brute force because $R=\max(P,Q)=P$ for all intervals. The interval length in these simulations is $\Delta=1000$.
}
\label{AdatomConvergence}
\end{figure}

For this system we also examine the pattern of convergence of the algorithm by
a random resampling procedure. For each value of $Q$ we collect data on the
estimated probability every $10^7$ iterations. We use
the average over all these data to obtain a reference value for the probability that
the red adatoms reach each other.  We randomly pick from the saved values to add to a
list, until the error in the logarithm of the average over the list compared to the
logarithm of the reference value is less than
$\epsilon$.  We count how many random samples are needed to achieve 
$\epsilon$ in 10000 independent trials and report the average values over a
range of $\epsilon$ (Fig.\ \ref{AdatomConvergence}).  This procedure lets us
determine the pattern of convergence from a single set of data. 

At small values of $\epsilon$, both 
brute force and STePS require an amount of time proportional to
$1/\epsilon$ (though the prefactor is generally smaller for STePS when using
reasonable values of $Q$). However, at large values of $\epsilon$, brute force
saturates to a constant time (effectively the time needed to observe
the first non-zero event), whereas STePS continues to decay as
$1/\epsilon$ or even more steeply. We find that for $\epsilon> 2$, 
the convergence curves for different $Q$ values can cross, suggesting
that larger $Q$ values may be better for getting rough estimates but that
smaller $Q$ values converge faster when more accuracy is desired.

To see how STePS performs on more challenging problems, we simulated a system
with a density of adatoms of $\rho=0.05$.  This problem is sufficiently
difficult that we do not have a brute force reference value. Instead we use a
long run at $Q=0.95$ and use the convergence curve resampling procedure described
above to estimate the time necessary to obtain the desired accuracy. For this
system, we estimate the probability that the red atoms reach each other to be
$(3.0\pm 1.4) \times 10^{-9}$  (using 3200 samples taken every
$10^{7}$ iterations). We expect this problem to be roughly $100$ times harder
for brute force than $\rho=0.03$.  When we measure the computational time
needed to obtain an error in the logarithm of $\epsilon=2$ for STePS, we find
that it is $2.8 \times 10^8$ iterations for $\rho=0.03$ and $3.2 \times 10^9$
iterations for $\rho=0.05$. As such, while the brute force simulation
is more difficult by a factor of $100$, the STePS simulation is more
difficult by only a factor of $10$ (combining this with the factor of 20
reported above for $\rho=0.03$, the speed-up relative to brute force is 200).
We expect this trend to continue as red adatom contact becomes rarer.

Finally, we demonstrate the use of STePS to compute a path-dependent quantity
other than the final probability. In this problem the successful trajectories
are distinct from those of the absorbing channel in that particles must not
only avoid certain collisions (with the walls of the channel or blue adatoms)
but also make others (with the other red adatom). Note that we do not reference
this need to make contact in the STePS progress constraint. Nevertheless, this
secondary requirement should favor shorter paths over longer ones, as the
likelihood of blue-adatom contact grows exponentially in time.  Consequently,
directed motion of the red adatoms should be favored.  When we do not specify the
requirement for red adatoms to make contact, only that they avoid blue adatoms,
we should recover ordinary diffusion.

\begin{figure}[t]
\includegraphics[width=\columnwidth]{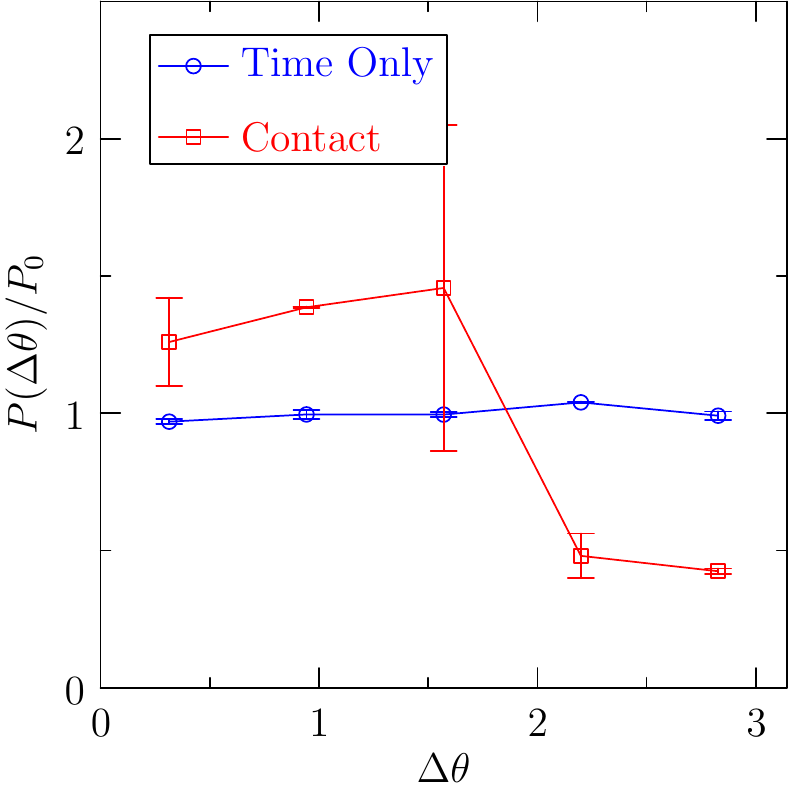}
\caption{Distribution of the change in direction of travel of the red adatoms along their trajectories at $\rho=0.05$.  We compare the case in which we look at all paths that survive for a fixed time interval ($4920$ iterations) to the case in which we select only paths that bring the red adatoms into contact. The data show is computed by averaging the upper-half-plane data (positive angles) with the lower-half-plane data (negative angles). The error bars are computed by taking the standard deviation of the two data sets from this mean. }
\label{FigCorrelation}
\end{figure}

We test this hypothesis by measuring the persistence of the direction of
motion. We take the paths generated by the algorithm and at each point in time
determine the direction of motion averaged over 1230 iterations, which corresponds
to approximately $10$ moves of one of the red adatoms. 
We then examine the distribution of differences between the
angle of travel at time $t$ and time $t+2460$ iterations.
For ordinary diffusion these motions
should be uncorrelated, giving rise to a flat distribution. If the collision
condition selects for paths in which the particles preferentially move towards
each other, this will show up as a peak in the angle distribution. We correct
for the discrete nature of motions on the lattice at short times by computing
the distribution $p(\Delta\theta)$ relative to the result when the density of
blue adatoms is zero ($p_0$). The results are in Fig.\ \ref{FigCorrelation}.
Using STePS, we are able to calculate this function at densities beyond
those accessible to brute force ($\rho=0.05$).  As predicted, when we remove
the requirement for contact, the distribution of change in travel direction
corresponds to ordinary diffusion, whereas when we look at paths in which the
red adatoms come into contact we obtain a peak in the distribution.

\section{Discussion}

Here, we introduced the STePS algorithm for steered transition path sampling
and demonstrated its use for several quantitative rare event problems.
Although we used overdamped Langevin dynamics as the basis for our
analytical arguments about the properties of the algorithm, it can be used with
any stochastic dynamics.  
Moreover, because integration is always forward in
time in the method, it can be used to study systems that are out of
equilibrium:  transient relaxation from an arbitrary initial condition and
dynamics that are intrinsically microscopically irreversible.  The output of
the method is a full transition path, from which multipoint time correlation
functions can be calculated directly.  

The progress constraint is most readily expressed in terms of whether an
elementary event does or does not happen within the integration interval,
although more complex specifications that span longer periods are possible.  As
such, the method is particularly well suited to controlling currents of dynamic
events.  At the same time, it is considerably simpler to implement than methods
that directly modify the dynamics for this purpose
\cite{Giardina2006,Tchernookov2010b} because it does not require a simple form
for the microscopic transition rates.

In the present manuscript, we illustrate the control of currents of 
dynamic events by suppressing
collisions.  In the highly diffusive systems considered (the absorbing channel
and the adatom surface), shooting full trajectories becomes inefficient because
the short correlations prevent transmission of information from one part of a
trajectory to another.  STePS, in contrast, allows accumulated progress to be
maintained even when an undesired event occurs.  This could be useful for
mechanisms with many intermediates, such as 
isomerizations in materials and certain biomolecular systems.  By the same token, we expect the method to outperform shooting
in enhancing or suppressing mobility in models of glassy dynamics, where it can
be necessary to have a coincidence of facilitating features
\cite{Merolle2005,Hedges2009};  it would be interesting to compare the method
with alternative path sampling schemes in which successive paths are correlated
by construction \cite{Crooks2001}.

The core concept of the method is that the dynamics are biased and then
reweighted for the calculation of averages.  It is thus natural to ask how the
method compares with dynamic importance sampling (DIMS)
\cite{Zuckerman2001,Perilla2011}.  We show in Appendix \ref{secn:drift} that
the progress constraint can be cast as an adaptive drift.  The essential
difference, however, is that the bias in STePS is applied only when needed.
The dynamics are not modified when the system naturally tends to make progress,
as when going downhill in free energy.  Moreover, when the dynamics are biased,
it is by reweighting trajectory segments that are generated with the original
dynamics.  Together, these features maximize the overlap with the original path
ensemble and limit the extent of reweighting.  It is conceivable that an
adaptive form of DIMS with these features could be developed.  
The optimal design of rare event simulation techniques is an active area of research.
A detailed analysis of efficient choices of biasing functions in DIMS can be found in \cite{Vanden-Eijnden2012};  
the optimal design of DIMS in the context of highly oscillatory energy landscapes has been addressed in  \cite{Dupuis2011}.
The conclusions drawn in those studies \cite{Vanden-Eijnden2012,Dupuis2011} may be of qualitative use in STePS applications as well. 

\appendix

\section{Error in estimates}
\label{secn:bias}

In this section, we describe how the results of the algorithm depend on
the procedure used to determine $P$, the estimated probability of satisfying
the progress constraint in an interval.  As discussed in the main text, we
generate trajectory segments in batches of size $M$ until at least one makes
progress.  We denote the total number of trajectory segments by $N$ and their
values at the end time interval of length $\Delta$ by
$\{x_j(\Delta)\}_1^N.$  The
numbers of segments that do and do not satisfy the progress constraint are
denoted by $N^+$
and $N^-$, respectively.  In general, we expect $N$, $N^+$ and $N^-$ to depend
on the point in phase space from which we are shooting, $x(0)$, but we do not
explicitly write $x(0)$ as an argument of these quantities below for clarity.  

Let the number of batches be the random integer $L$, such that $N = LM$.  The
fact that the quantity $N$ fluctuates creates a small systematic error in $P$.
To see this, denote the true probability of satisfying the progress constraint
as 
\begin{equation}
p_s = \mathbf{P} \left( \phi(x(\Delta))\geq \phi(x(0))\right),
\end{equation}
where $\mathbf{P}$ denotes the probability of an event at point $x$, 
$\phi$ is the order parameter defining the progress constraint.

The expectation ($\mathbf{E}$) of $P = N^+/N$ can be written as an average over
all possible values of $L$ of a sum over an indicator function ($h = 1$ when the argument condition is fulfilled, and $h=0$ otherwise) that selects for exactly $L$ batches multiplied by the number of successes in the last
batch  divided by $N = LM$:
\begin{widetext}
\begin{align}
\mathbf{E}\left[ \frac{N^+}{N}\right] 
&= \mathbf{E}\left[\sum_{l=1}^\infty \frac{1}{lM}h\left(L = l\right)
\sum_{j=1}^Mh\left( \phi(x_{(l-1)M+j}(\Delta))\geq \phi(x(0))\right) \right] 
\notag \\
&=  \frac{1}{M} \sum_{l=1}^\infty \sum_{j=1}^M \frac{1}{l}\mathbf{P}\left(L =
l
 \text{ and } \phi(x_{(l-1)M+j}(\Delta))\geq  \phi(x(0)) \right) 
\notag \\
& =  \frac{1}{M} \sum_{l=1}^\infty \sum_{j=1}^M 
\frac{1}{l}\mathbf{P}\left(\phi(x_k(\Delta))<\phi(x(0))\text{ for all
}k\leq(l-1)M \text{ and } \phi(x_{(l-1)M+j}(\Delta))\geq  \phi(x(0)) \right)
\notag \\
& =  p_s\sum_{l=1}^\infty \frac{1}{l}(1-p_s)^{(l-1)M} \notag \\
& =p_s \left(-\frac{\ln( 1- (1-p_s)^M)}{(1-p_s)^M} \right) \notag\\
&= p_s \left(1+\mathcal{O}\left((1-p_s)^M\right)\right).\label{expNplus}
\end{align}
Thus $\mathbf{E}[N^+/N]$ differs from $p_s$;
no such systematic error arises if $N$ is fixed.
Of course, estimating $p_s$ is not our overall goal.  However, for any reasonable test function $f$,
\begin{multline}\label{expfs3}
\mathbf{E}\left[f(\hat{x}(\Delta))W| N^+, N^-\right] = 
\mathbf{E}\left[f(x(\Delta))| \phi(x(\Delta))\geq\phi(x(0))\right]
(N^+/N) \\
+\mathbf{E}\left[f(x(\Delta))
| \phi(x(\Delta))<\phi(x(0))\right] (N^-/N),
\end{multline}
where $\hat{x}(\Delta)$ is the selected position at the end of the interval, and $W$ is the weight for the trajectory segment.
Taking the expectation over $N^+$ and $N^-$ we obtain
\begin{multline}\label{expfs4}
\mathbf{E}\left[f(\hat{x}(\Delta))W\right] =  \mathbf{E}\left[f(x(\Delta))|
\phi(x(\Delta))\geq\phi(x(0))\right]
p_s \left(1+\mathcal{O}\left((1-p_s)^M\right)\right)  \\
+\mathbf{E}\left[f(x(\Delta))
| \phi(x(\Delta))<\phi(x(0))\right] \left(1-
p_s\left(1+\mathcal{O}\left((1-p_s)^M\right)\right) \right)
\end{multline}
\end{widetext}
or, after combining terms,
\begin{multline}\label{expfs5}
\mathbf{E}\left[f(\hat{x}(\Delta))W\right] = 
\mathbf{E}\left[f(x(\Delta))\right]\\
+ \mathcal{O}\left(p_s (1-p_s)^M\right),
\end{multline}
i.e., in one interval of the scheme a small systematic error of $\mathcal{O}\left(p_s
(1-p_s)^M\right)$ is incurred, but it is fairly easy to control.  One can
increase $M$ or modify $\Delta$ to increase $p_s$.  In practice, we
try to learn $p_s$ from the simulation and then choose $M= c/p_s$, in which case
the error decreases uniformly exponentially in the constant $c$.  By iterating
the argument above we can see that over finitely many time steps the error
remains small.  This is consistent with our numerical experience.

\section{Continuous time limit}
\label{secn:drift}

It is of interest to consider the dynamics to which the STePS algorithm tends as $\Delta \rightarrow 0$ to set the method in the context of others
that bias the dynamics \cite{Zuckerman2001,Perilla2011} and further understand how best to choose its parameters.  For definiteness and simplicity we base our discussion on overdamped Langevin dynamics (Eq.\ \eqref{eq:overdamped}).
A similar set of manipulations for inertial Langevin dynamics (which we leave to
the reader)
suggests that in an inertial context, for small $\Delta,$ we should choose a
$\phi$ that depends on velocity.
Our notation is the same as in the previous section.  Starting from position
$x$, we simulate $N$ independent samples 
$\{x_j(\Delta)\}_1^N$  of $x(\Delta)$ and divide them into  $N^+$ satisfying 
$\phi(x_j(\Delta))\geq \phi(x(0))$ and $N^-$ satisfying $\phi(x_j(\Delta))<
\phi(x(0)).$  Next we select between the two groups with a  probability depending
only on the ratio $N^+/N$, which we denote by $R(N^+/N).$  Finally we pick
$\hat{x}(\Delta)$ uniformly from the selected group.  To simplify the formulas
in this section we assume that $N$ is 
a deterministic function of position and is not random.  

\begin{widetext}
Straightforward manipulations reveal that, for any reasonable function $f$, 
  \begin{multline}\label{expf4}
\mathbf{E}\left[ f(\hat{x}(\Delta))\right]
 =  \mathbf{E}\left[f(x_1(\Delta))\right]\\
+ \mathbf{E}\left[R\left(\frac{N^+}{N}\right) - \frac{N^+}{N}\right]
\bigg(   \mathbf{E}\left[ f(x_1(\Delta))| \phi(x_1(\Delta))\geq\phi(x(0))\right]
-
\mathbf{E}\left[ f(x_1(\Delta))| \phi(x_1(\Delta))<\phi(x(0))\right]\bigg).
\end{multline}
This formula can be used to compute, for example, the mean displacement and displacement squared over one interval. Indeed, substituting $f(y)=y$ into Eq.\
\eqref{expf4} and
 recalling that for small $\Delta$
\begin{equation}
x(\Delta) \approx x(0) + F(x(0))\Delta + \sqrt{2T} B(\Delta),
\end{equation}
we obtain (to leading order in each term) a mean displacement of
\begin{multline}
\mathbf{E}\left[ \hat{x}(\Delta)\right] - x(0)
 \approx  F(x(0))\Delta \\
+ \sqrt{2T}\mathbf{E}\left[R\left(\frac{N^+}{N}\right) - \frac{N^+}{N}\right]
\bigg(   \mathbf{E}\left[B(\Delta)| \langle B(\Delta), \nabla
\phi(x(0))\rangle\geq 0\right] -
\mathbf{E}\left[B(\Delta)| \langle B(\Delta), \nabla \phi(x(0))\rangle< 0
\right]\bigg).
\end{multline}
In this case, the expectations over $B(\Delta)$ can be computed exactly to yield
   \begin{equation}\label{mnd}
   \mathbf{E}\left[ \hat{x}(\Delta)\right] - x(0)
 \approx   +F(x(0))\Delta 
+4\sqrt{\frac{T\Delta}{\pi}} \mathbf{E}\left[R\left(\frac{N^+}{N}\right) -
\frac{N^+}{N}\right] 
 \frac{\nabla \phi(x(0))}{\lVert  \nabla \phi(x(0))\rVert}.
\end{equation}
\end{widetext}
Similarly, substituting $f(y) = (y-x(0))^2$ one can compute to leading order
   \begin{equation}\label{msd}
   \mathbf{E}\left[ \left(\hat{x}(\Delta)-x(0)\right)^2\right] 
 \approx   2T\Delta.
\end{equation}

Eq.\ \eqref{mnd} shows that, to obtain a meaningful limiting process as
$\Delta\rightarrow 0,$ the second term on the right hand side must be of order
$\Delta.$  Choosing the function $R$ to be of the form
 \begin{equation}\label{r}
R\left(p \right) = p + r\left( p\right)\sqrt{\Delta} 
 \end{equation}
 for some function $r$ achieves this goal.  
 Our expression in Eq.\ \eqref{r} is of practical use.   Note that $R(N^+/N) =
N^+/N$ is the natural probability of selecting the group of $x_j(\Delta)$ with
$\phi(x_j(\Delta))>\phi(x(0))$ in the sense that it results in a process identical
in law to the original underlying process and in weights equal to one.  Thus Eq.\
\eqref{r} suggests that for small values of $\Delta$ one should perturb this
natural selection probability by no more or less than an amount of order
$\sqrt{\Delta}.$
 
To obtain the precise form of the limiting dynamics, define the function
\begin{equation}
 b = F + 4\sqrt{\frac{T}{\pi}} \frac{r\nabla \phi}{\lVert \nabla \phi \rVert}.
\end{equation}
Eqs.\ \eqref{mnd} and \eqref{msd} with $R$ of the form in Eq.\ \eqref{r} suggest that,
 as $\Delta\rightarrow 0,$ STePS converges to the solution $\bar x$ of the stochastic differential equation
\begin{equation}\label{ybar}
\dot{\bar x}(t) = b(\bar x(t))dt + \sqrt{2T} \dot{\bar B}(t)
\end{equation}
where $\bar B$ is a Brownian motion (but not the one driving the
underlying system).
In fact this can be confirmed using, for example, Theorem 4.1 in Chapter 7 of
\cite{Kurtz2005}.  The fact that STePS approaches a well defined limiting
process as $\Delta\rightarrow 0$ is of little comfort if the weights $W$ are
not also well behaved in this limit.  Fortunately this too can be established.
We note, however, that the limit cannot be expressed in terms of the path of
$\bar x$  in Eq.\ \eqref{ybar}.
A similar phenomenon occurs when Eq.\ \eqref{eq:overdamped} is represented
by the simple weakly accurate discretization  
\begin{multline}
x_{n+1} = x_n + F(x_n)\Delta \\ + \sqrt{2T\Delta }\text{sgn}\left(B((n+1)\Delta)-B(n\Delta)\right).
\end{multline}
In fact, if one biases the above process by replacing the sign of the Brownian increments by a variable $\xi_n$ with
\begin{equation}
\mathbf{P}\left( \xi_n = \pm 1\right) =  \frac{1}{2} \pm  r\left( x_n\right)\sqrt{\Delta}
\end{equation}
and reweights accordingly, then the behavior of the resulting weights as $\Delta\rightarrow 0$ is completely analogous to the limiting behavior of the STePS weights.

\begin{table}
\begin{displaymath}
\begin{array}{lcc} \hline
\Delta &   \mathbf{E}[\hat{x}]        &  \mathbf{E}[\hat{x}^2]          \\ \hline
1     &  0.08\pm 0.02 &  1.03\pm 0.02    \\
0.1   &  0.08\pm 0.01 &  1.08\pm 0.02    \\
0.01  &  0.08\pm 0.02 &  1.00\pm 0.02    \\
0.001 &  0.07\pm 0.03 &  0.99\pm 0.01    \\ \hline
\end{array}
\end{displaymath}
\caption{\label{tab:msd} Means of the displacement and displacement squared
at time 1 for STePS on an overdamped Langevin dynamics with $F = 0$ and $\sqrt{2T} = 1$; $r = 0.05$ and $N = 10$ in the
protocol employed in Appendix \ref{secn:drift}.  The predicted values for
$\mathbf{E}[\hat{x}]$ and $\mathbf{E}[\hat{x}^2]$ are
$2r\sqrt{2/\pi}$ and 1, respectively.}
\end{table}

\begin{table}
\begin{displaymath}
\begin{array}{lcc} \hline
\Delta & \mathbf{E}[W]             &\mathbf{E}[W^2] \\ \hline
1      & 0.999\pm 0.002   & 1.008\pm 0.003 \\
0.1    & 1.002\pm 0.001 & 1.014\pm 0.001 \\
0.01   & 0.999\pm 0.001   & 1.009\pm 0.002 \\
0.001  & 0.994\pm 0.003   & 0.999\pm 0.007  \\ \hline
\end{array}
\end{displaymath}
\caption{\label{tab:weight} Statistics of the path weights for the simulations in 
Table \ref{secn:drift}.}
\end{table}

Tables \ref{tab:msd} and \ref{tab:weight} illustrate the limiting behaviors of
the motion and the overall path weights for the simple case of $F = 0$ and
$\sqrt{2T} = 1$.  We see that the means of the displacement and displacement
squared tend to their predicted values, and the weights remain well behaved as
$\Delta$ becomes small.  Thus STePS tends to a dynamics similar in form to DIMS
\cite{Zuckerman2001,Perilla2011}, but we caution the reader from
over-interpreting this result.  The goal of STePS is not to approximate a
continuous time importance sampling method, and the adaptivity of STePS depends
crucially on its discrete time structure.

\begin{acknowledgments}
N.G. was supported by NSF DMR-MRSEC 0820054.
\vfill
\end{acknowledgments}

\bibliographystyle{apsrev}
\bibliography{steps}

\end{document}